\documentclass[twocolumn,superscriptaddress,amsmath, amssymb, amsfonts,preprintnumbers,aps,prd,longbibliography,nofootinbib]{revtex4-1}

\renewcommand{\sec}[1]{\textit{#1. --- }}
\usepackage{graphicx}% Include figure files
\usepackage{dcolumn}% Align table columns on decimal point
\usepackage{bm}% bold math
 \usepackage{amstext}
 \usepackage{amssymb}
\usepackage{subcaption}
 \usepackage{amsmath}
 \usepackage{graphicx}
 \usepackage{color}
 \usepackage{bbold}
 \usepackage{delimset} % for brackets
\usepackage[colorlinks=true]{hyperref}
\usepackage{orcidlink}

\usepackage{tikz}
\usetikzlibrary{decorations.pathmorphing}
\usetikzlibrary{decorations.markings}
\usetikzlibrary{positioning, shapes, snakes, arrows}
 \usepackage{tikz-feynman}

\tikzset{ 
	graviton/.style={line width=.8pt, -latex,decorate, decoration={snake, segment length=4pt,amplitude=1.8pt, pre length=.1cm, post length=.25cm}},
	worldline/.style={gray, line width=1pt},
	worldlineBold/.style={black, line width=.6pt},
        background/.style={black,dotted,line width=1pt},
	zUndirected/.style={line width=1pt},
	zParticle/.style={line width=1pt,postaction={decorate},decoration={markings,mark=at position .6 with {\arrow[#1]{latex}}}},
	zParticleF/.style={line width=1pt,postaction={decorate}},
	cscalar/.style={line width=1pt,postaction={decorate},decoration={markings,mark=at position .6 with {\arrow[#1]{latex}}}},
	cscalar2/.style={line width=1pt,postaction={decorate},decoration={markings,mark=at position .8 with {\arrow[#1]{latex}}}},
	photon/.style={line width =.8pt, decorate, decoration={snake, segment length=3pt, amplitude=1.8pt,  pre length=.1cm, post length=.1cm}},
	 mid arrow/.style={postaction={decorate,decoration={
        markings,
        mark=at position .5 with {\arrow[#1]{latex}}}}} ,
        worlddot/.style={dotted, line width=.8pt},
	worlddot2/.style={dotted, line width=1pt}   }

\DeclareFontFamily{OT1}{pzc}{}
\DeclareFontShape{OT1}{pzc}{m}{it}{<-> s * [1.350] pzcmi7t}{}
\DeclareMathAlphabet{\mathpzc}{OT1}{pzc}{m}{it}

\def\cN{\mathcal{N}}
\def\cO{\mathcal{O}}

\def\eps{\epsilon}

\def\d{\mathrm{d}}
\def\D{\mathrm{D}}
\def\eE{\mathrm{E}}
\def\eK{\mathrm{K}}

\def\dd{\delta}

\def\d{\mathrm{d}}
\def\eps{\epsilon}

\def\braket#1{\langle #1 \rangle}

\def\nn{\nonumber}

\def\Eqn#1{Eq.~\eqref{#1}}

\def\Fig#1{Fig.~{\ref{#1}}}

\def\Rcite#1{Ref.~\cite{#1}}

\newcommand*\Bell{\ensuremath{\boldsymbol\ell}}

\newcommand{\vev}[1]{\langle #1\rangle}

\newcommand{\be}{\begin{equation}}
\newcommand{\ee}{\end{equation}}
\newcommand{\ba}{\begin{align}}
\newcommand{\ea}{\end{align}}
\ifx\genfrac\sdflkaj\else\fi
\newcommand{\sfrac}[2]{{\textstyle\frac{#1}{#2}}}

\newcommand{\mn}{{\mu\nu}}

\newcommand{\pat}{\partial}

\begin{document}

\preprint{HU-EP-23/16-RTG}

\title{Conservative scattering of spinning black holes at fourth post-Minkowskian order 
}

\author{Gustav Uhre Jakobsen\,\orcidlink{0000-0001-9743-0442}} 
\email{gustav.uhre.jakobsen@physik.hu-berlin.de}
\affiliation{%
Institut f\"ur Physik und IRIS Adlershof, Humboldt Universit\"at zu Berlin,
Zum Gro{\ss}en Windkanal 2, 12489 Berlin, Germany
}
 \affiliation{Max Planck Institute for Gravitational Physics (Albert Einstein Institute), Am M\"uhlenberg 1, 14476 Potsdam, Germany}

\author{Gustav Mogull\,\orcidlink{0000-0003-3070-5717}}
\email{gustav.mogull@aei.mpg.de} 
\affiliation{%
Institut f\"ur Physik und IRIS Adlershof, Humboldt Universit\"at zu Berlin,
Zum Gro{\ss}en Windkanal 2, 12489 Berlin, Germany
}
 \affiliation{Max Planck Institute for Gravitational Physics (Albert Einstein Institute), Am M\"uhlenberg 1, 14476 Potsdam, Germany}
 
 \author{Jan Plefka\,\orcidlink{0000-0003-2883-7825}} 
\email{jan.plefka@hu-berlin.de}
\affiliation{%
Institut f\"ur Physik und IRIS Adlershof, Humboldt Universit\"at zu Berlin,
Zum Gro{\ss}en Windkanal 2, 12489 Berlin, Germany
}

\author{Benjamin Sauer\,\orcidlink{0000-0002-2071-257X}} 
\email{benjamin.sauer@hu-berlin.de}
\affiliation{%
Institut f\"ur Physik und IRIS Adlershof, Humboldt Universit\"at zu Berlin,
Zum Gro{\ss}en Windkanal 2, 12489 Berlin, Germany
}

\author{Yingxuan Xu\,\orcidlink{0000-0001-6135-8864}} 
\email{yingxu@physik.hu-berlin.de}
\affiliation{%
Institut f\"ur Physik, Humboldt Universit\"at zu Berlin,
Newtonstra{\ss}e 15, 12489 Berlin, Germany
}

\begin{abstract}
Using the $\cN=1$ supersymmetric, spinning worldline quantum field theory formalism
we compute the conservative spin-orbit part of the 
momentum impulse $\Delta p_i^\mu$, spin kick $\Delta S_i^\mu$ and scattering angle $\theta$
from the scattering of two spinning massive bodies
(black holes or neutron stars)
up to  fourth post-Minkowskian (PM) order. These three-loop results
extend the state-of-the-art for generically spinning binaries from 3PM to 4PM. They are 
obtained by employing recursion relations for the integrand construction and advanced multi-loop Feynman integral technology
in the causal (in-in) worldline quantum field theory framework to directly produce classical observables.  
We focus on the conservative contribution (including tail effects) 
and outline the computations for the dissipative contributions as well.
Our spin-orbit results agree with N$^3$LO post-Newtonian and test-body 
data in the respective limits. We also re-confirm the conservative 4PM non-spinning results.
\end{abstract}
 
\maketitle 

High-precision predictions for the gravitational waves emitted from the interaction of
compact binaries are essential for data analysis of gravitational wave detectors
\cite{Abbott:2016blz,LIGOScientific:2017vwq,LIGOScientific:2018mvr,LIGOScientific:2020ibl,LIGOScientific:2021usb,LIGOScientific:2021djp}. They are the prerequisites to address fundamental questions in astro-, gravitational, 
particle and nuclear physics through observations of gravitational waves. 
The third generation of detectors --- LISA, Einstein Telescope and Cosmic Explorer~\cite{Ballmer:2022uxx},
scheduled to go online in the 2030s ---
will reach an experimental accuracy that goes well beyond the present state-of-the-art in 
analytical and numerical gravitational wave physics \cite{Purrer:2019jcp,Kalogera:2021bya}. 
This situation has sparked a renewed effort 
to extend and innovate traditional approaches to the classical relativistic two-body problem.

The early inspiral phase of a bound two-body system, or a small-deflection scattering scenario,
is characterized by a scale separation between the relative distance of the compact bodies and their sizes. 
Here, the weakness of the gravitational field enables an analytic, perturbative treatment:
one models black holes (BHs) or neutron stars (NSs) as two massive, spinning
point particles that interact gravitationally, controlled by a perturbative expansion in
Newton's coupling $G$ \cite{Westpfahl:1979gu,Bel:1981be,Damour:2017zjx}. In addition, finite-size and tidal effects may be included in the point-particle model by coupling higher-dimensional operators to the particle's worldline theory
in the logic of effective field theory (EFT)
\cite{Goldberger:2004jt,Porto:2016pyg,Levi:2018nxp}. Typically for bound orbits this is done in a post-Newtonian (PN)
expansion in \emph{both} $G$ and the relative velocity $v$ of the bodies, both
in the traditional \cite{Blanchet:2013haa,Schafer:2018kuf,Futamase:2007zz,Pati:2000vt,Bel:1981be} and 
effective field theory 
\cite{Goldberger:2004jt,Goldberger:2006bd,Goldberger:2009qd,Kol:2007bc,Goldberger:2007hy,Foffa:2013qca,Rothstein:2014sra,Porto:2016pyg,Levi:2018nxp,Galley:2009px,Foffa:2019hrb,Blumlein:2020pyo,Bini:2020wpo,Bini:2020hmy}
based approaches. The present state of the art is approaching the 5PN level \cite{Blumlein:2020znm,Bini:2020nsb,Bini:2020wpo,Bini:2020hmy,Bini:2020uiq,Blumlein:2021txe,Foffa:2019eeb,Almeida:2023yia}
including N$^3$LO spin effects \cite{Levi:2022rrq, Kim:2022bwv,Kim:2022pou,Mandal:2022ufb,Mandal:2022nty,Levi:2020uwu,Levi:2020kvb}.

However, quantum field theory (QFT) based methods using the post-Minkowskian (PM)
expansion \cite{Bertotti:1956pxu,Kerr:1959zlt,Portilla:1980uz,Westpfahl:1985tsl}, i.e.~perturbative in $G$ but exact in $v$,  are rapidly developing 
(see \cite{Kosower:2022yvp,Bjerrum-Bohr:2022blt,Buonanno:2022pgc} for reviews):
They derive from the well-studied perturbative
quantization of Einstein's theory of gravity about flat space-time backgrounds.
Here, state-of-the-art Feynman integral technology
may be fruitfully ported to the realm of classical general relativity: 
integration by parts methods (IBP) for a reduction to  
master integrals \cite{Laporta:2000dsw,Smirnov:2008iw,Gehrmann:1999as},
their computation via differential equations \cite{Gehrmann:1999as,Henn:2013pwa} and the method of regions to determine
boundary values \cite{Beneke:1997zp}. Clearly, the natural habitat for the
PM expansion is the scattering scenario. Still, the scattering data in the conservative sector
informs also the (conservative) bound case through a matching to a real
\cite{Cheung:2018wkq,Bjerrum-Bohr:2018xdl,Cristofoli:2019neg,Bern:2020buy,Kosmopoulos:2021zoq,FebresCordero:2022jts,Jakobsen:2022zsx} or effective-one-body
\cite{Buonanno:1998gg,Buonanno:2000ef,Antonelli:2019ytb,Damgaard:2021rnk,Khalil:2022ylj} Hamiltonian.

At present two complementary QFT-based approaches to the PM expansion are being pursued:
based on scattering amplitudes~\cite{Iwasaki:1971vb,Holstein:2004dn,Neill:2013wsa,Luna:2017dtq,Bjerrum-Bohr:2013bxa,Bjerrum-Bohr:2018xdl,Bern:2019nnu,Bern:2019crd,Bjerrum-Bohr:2021wwt,Cheung:2020gyp,Bjerrum-Bohr:2021din,DiVecchia:2020ymx,DiVecchia:2021bdo,DiVecchia:2022piu,Damour:2020tta,Herrmann:2021tct,Damgaard:2019lfh,Brandhuber:2021eyq}
and worldlines~\cite{Kalin:2020mvi,Kalin:2020fhe,Kalin:2020lmz,Dlapa:2021npj,Dlapa:2021vgp,Mougiakakos:2021ckm,Riva:2021vnj,Mougiakakos:2022sic,Riva:2022fru,Mogull:2020sak,Jakobsen:2021smu,Jakobsen:2021lvp,Jakobsen:2021zvh,Jakobsen:2022fcj,Jakobsen:2022psy}.
In the scattering scenario there are three key observables: the deflection of momentum
(known as impulse),
the change of spin vectors (known as spin kick) and the Bremsstrahlung (waveform)
in the far-field limit. Ignoring finite-size, tidal and spin effects the impulse is known at 3PM (two-loop) 
\cite{Bern:2019nnu,Bern:2019crd,Cheung:2020gyp,Kalin:2020fhe,Kalin:2020lmz,Brandhuber:2021eyq} and 4PM (three-loop)
\cite{Bern:2021dqo,Bern:2021yeh,Dlapa:2021npj,Dlapa:2021vgp,Dlapa:2022lmu,Dlapa:2023hsl,Bjerrum-Bohr:2022ows} order.
Preliminary work has also begun on the 5PM scattering angle, using electrodynamics as a toy model~\cite{Bern:2023ccb}.
In the case of spinning binaries the impulse is known up to quintic spin interactions \cite{Bern:2022kto,Aoude:2022thd} at 2PM order,
and up to quadratic spin interactions (including the spin
kick) at 3PM order \cite{Jakobsen:2022fcj,Jakobsen:2022zsx}\cite{FebresCordero:2022jts} together with partial results at all spin orders \cite{Vines:2017hyw,Alessio:2022kwv,Aoude:2022thd,Aoude:2023vdk,Alessio:2023kgf}.
Leading tidal effects have been computed at 2PM \cite{Bini:2020flp,Haddad:2020que}
and 3PM order \cite{Cheung:2020sdj,Kalin:2020lmz,Jakobsen:2022psy}.
For the Bremsstrahlung waveform the leading-order
result without \cite{1975Kovacs,1977CrowleyThorne,Kovacs:1977uw,Kovacs:1978eu,Jakobsen:2021smu,Mougiakakos:2021ckm} and with spin \cite{Jakobsen:2021lvp,Riva:2022fru} or tidal effects
\cite{Mougiakakos:2022sic,Jakobsen:2022psy} was recently updated to 
next-to-leading order for the non-spinning case \cite{Brandhuber:2023hhy,Herderschee:2023fxh,Georgoudis:2023lgf,Elkhidir:2023dco}.

In this Letter we provide the conservative, spin-orbit contributions to the impulse and spin kick at 4PM accuracy,
together with the total scattering angle.
These results provide the basis to refine effective one-body Hamiltonians and resummed scattering prescriptions for high-precision gravitational wave physics.
Our worldline quantum field theory (WQFT) hinges on three innovations to the EFT approach for gravitational scattering:
(i) quantizing \emph{both} the worldline degrees of freedom and the gravitational field allows for a
diagrammatic formulation of the classical perturbation theory yielding the observables 
as one-point functions of the worldline or gravitational fields \cite{Mogull:2020sak},
(ii) capturing the spin of the compact
objects through a supersymmetric worldline theory~\cite{Jakobsen:2021zvh},
(iii) the Schwinger-Keldysh (in-in) initial value formulation of WQFT that induces
the use of retarded propagators and a causality flow in the diagrammatic expansion \cite{Jakobsen:2022psy}.

\sec{Supersymmetric in-in WQFT formalism}
The effective worldline theory of spinning bodies (Kerr BHs or NSs) with masses $m_i$ and space-time coordinates $x_i^\mu(\tau)$ on a general $D$-dimensional space-time with metric $g_{\mu\nu}$ is described up to quadratic order in spin by an $\cN=2$ supersymmetric worldline theory~\cite{Jakobsen:2021zvh}.
As we are focusing on the spin-orbit (linear-in-spin) dynamics here, the $\cN=1$ incarnation of this theory will suffice:
\begin{align}\label{eq:action}
S=-\sum_{i=1}^{2}m_{i}\!\int\!\d\tau\bigg[\sfrac{1}{2}g_{\mu\nu}\dot x_i^{\mu}\dot x_i^{\nu}
	\!+\! i\psi_{i,a}\!\frac{\D\psi_i^a}{\D\tau}\!\bigg] + S_{\rm EH}\,.
\end{align}
The real anti-commuting vectors $\psi_i^a(\tau)$ are defined in a flat tangent space 
using the vierbein $e^{\mu}_{a}$
and $\frac{\D\psi_i^a}{\D\tau}=\dot\psi_i^a+\dot x^\mu{\omega_\mu}^{ab}\psi_{i,b}$
with the spin-connection $\omega _{\mu}{}^{ab}$ (our metric is mostly minus).
We work in $D=4-2\eps$ dimensions with $S_{\rm EH}$ the bulk Einstein-Hilbert action
including a gauge-fixing term;
the process of dimensional regularisation, wherein we ultimately send $\eps\to0$,
is aided by only this part of the full action needing to be lifted to $D$ dimensions.
The  $\psi_i^a(\tau)$ carry the spin degrees of freedom with the
spin tensors 
$S_i^{\mu\nu}=-i m_i \psi_i^{\mu}\psi_i^{\nu}$
and the Pauli-Lubanski vectors $S^\mu_i=m_i a_i^{\mu}=\frac12 \epsilon^{\mu}_{\,\nu\rho\sigma} v_i^\nu S_i^{\rho\sigma}$.

We expand the fields around their respective backgrounds:
the metric $g_\mn=\eta_\mn + \kappa h_\mn$, with $\kappa=\sqrt{32\pi G}$, and the worldlines
\begin{align}
\begin{aligned}\label{backgroundexp}
	x_i^\mu(\tau) &= b_i^\mu \!+\! v_i^\mu \tau \!+\! z_i^\mu(\tau)\,, &
	\psi^\mu_i(\tau) &= \Psi^\mu_i\!+\!{\psi'}_i^\mu(\tau)\,,
\end{aligned}
\end{align}
where $\{b_i^\mu,v_{i}^{\mu},\Psi_i^\mu\}$ are the initial (background) parameters of the two bodies.
Using background symmetries we set $b\cdot v_i=\Psi_i\cdot v_i=0$
where $b^\mu=|b|\,\hat b^\mu=b_2^\mu-b_1^\mu$ is the covariant impact parameter.
We also introduce the Lorentz factor $\gamma=v_1\cdot v_2$ and the relative velocity $v=\sqrt{\gamma^2-1}/\gamma$.

Causal observables including radiative effects arise from the Schwinger-Keldysh (in-in) formalism applied to WQFT \cite{Jakobsen:2022psy} where one doubles the fields: $h_{\mu\nu}\to(h^{(1)}_{\mu\nu},h^{(2)}_{\mu\nu})$ and  $Z_{i}^{\mu}\to(Z^{(1) \mu}_{i},Z^{(2) \mu}_{i})$ introducing the worldline ``super-fields'' $Z_{i}=\{z_{i},\psi'_{i}\}$.
Causal one-point functions follow from the in-in path integral
\begin{align}\label{ZWQFTdef}
\begin{aligned}
	\braket{\cO}:=\!\!
	\int\!{\cal D}[h^{(1,2)}_{\mu\nu},Z_i^{(1,2) \mu}]
	e^{i ( S[\{\, \}^{(1)}] - S[\{\, \}^{(2)}]^{\ast} )}\cO \, ,
\end{aligned}
\end{align}
normalized such that $\braket{1}=1$ and with $\{\,\}^{(n)}$ denoting the $(n)$'th copy of the doubled fields.
The key property we exploit is that the WQFT \emph{tree-level one-point functions} $\vev{Z^{(n)}_{i}}$ solve the classical equations of motion.
Moreover, the computation of one-point functions of in-in WQFT  reduces to the use of \emph{retarded} propagators combined with the standard in-out WQFT Feynman rules \cite{Jakobsen:2022psy}.
This formalism yields an efficient QFT-based scheme to solve the classical equations perturbatively.

Conservative observable can in turn be defined by neglecting all interactions between $h_\mn^{(1)}$ and $h_\mn^{(2)}$.
This may be achieved by using the in-in formalism only for the worldlines while keeping the in-out formalism for the gravitons and projecting on the real part of observables~\cite{Galley:2012hx,Kalin:2022hph}.
This separation of conservative effects at 4PM has proven its efficiency for the non-spinning results~\cite{Bern:2021yeh,Dlapa:2021vgp}.

\sec{WQFT Feynman rules}
The graviton propagator in de Donder gauge with Feynman prescription reads
\begin{align}\label{eq:gravProp}
	\begin{tikzpicture}[baseline={(current bounding box.center)}]
	\begin{feynman}
	\coordinate (x) at (-.7,0);
	\coordinate (y) at (0.5,0);
	\draw [photon] (x) -- (y) node [midway, below] {$k$} ;
	\draw [fill] (x) circle (.08) node [above] {$\mu\nu$};
	\draw [fill] (y) circle (.08) node [above] {$\rho\sigma$};
	\end{feynman}
	\end{tikzpicture}
        &=\frac{iP_{\mu\nu;\rho\sigma}}{k^{2}+i 0^{+}}\,,
\end{align}
with $P_{\mu\nu;\rho\sigma}:=\eta_{\mu(\rho}\eta_{\sigma)\nu}-\sfrac1{D-2}\eta_{\mu\nu}\eta_{\rho\sigma}$ while the worldline propagators associated with $z_i^\mu$ and $\psi_i^{\prime\mu}$ read, respectively
\begin{align}
  \begin{tikzpicture}[baseline={(current bounding box.center)}]
    \coordinate (in) at (-0.6,0);
    \coordinate (out) at (1.4,0);
    \coordinate (x) at (-.2,0);
    \coordinate (y) at (1.0,0);
    \draw [zUndirected] (x) -- (y) node [midway, below] {$\omega,n$} node [midway, above] {$\rightarrow$};
    \draw [background] (in) -- (x);
    \draw [background] (y) -- (out);
    \draw [fill] (x) circle (.08) node [above] {$\mu$};
    \draw [fill] (y) circle (.08) node [above] {$\nu$};
  \end{tikzpicture}&=\frac{-i\eta^{\mu\nu}}{m_i(\omega+i0^+)^{n}}\,
  \begin{cases}
 \text{$n=2$ for $z_{i}^{\mu}\,,$}\\  \text{$n=1$ for $\psi_{i}^{\prime\mu}\,.$}
  \end{cases} 
\end{align}
The arrow on the propagators indicates the momentum or energy flow on the \emph{retarded}
propagators.
Importantly, the Feynman graviton propagators reflect our focus on conservative observables.
Full dissipative results may be obtained by using retarded propagators instead.
The Feynman vertices of the spinning WQFT to lower multiplicities
have been exposed in \cite{Jakobsen:2021zvh}. 
The generic worldline vertex couples $n$ gravitons to $m$ worldline fields and reads
\begin{align}
\label{eq:WLvertexrule}
\raisebox{-20pt}{$V_{n|m} =$\!\!\!
\resizebox{0.15\textwidth}{!}{
\begin{tikzpicture}[baseline={(current bounding box.center)}]
    \coordinate (in) at (-1,0);
    \coordinate (out1) at (1,0);
    \coordinate (out2) at (1,0.5);
    \coordinate (out3) at (1,0.9);
    \coordinate (x) at (0,0);
      \node (k1) at (-.9,-1.3) {$k_1$};
    \node (k2) at (.9,-1.3) {$k_n$};
    \draw (out1) node [right] {$\omega_1$};
    \draw (out2) node [right] {$\!\!\!\vdots$};
     \draw (-.2,-0.8) node [right] {$\!\!\!\ldots$};
    \draw (out3) node [right] {$\omega_m$};
    \draw [background] (in) -- (x);
    \draw [zUndirected] (x) -- (out1);
    \draw [zUndirected] (x) to[out=30,in=180] (out3);
    \draw [photon] (x) -- (k1);
      \draw [photon] (x) -- (k2);
    \draw [fill] (x) circle (.08);
    \end{tikzpicture}$\sim$ }}& m \kappa^{n}\, e^{i k\cdot b} \dd\bigg(k\cdot 
    v+\sum_{i=1}^n\omega_i\bigg) \times \nn\\[-35pt] & \times 
   \begin{pmatrix} \text{polynomial in $\omega_{i},k_{j}$}\\
   \text{of degree $2n+m$}
   \end{pmatrix}\\[-5pt]\nn
\end{align}
where $k^{\mu}=\sum_{i=1}^{n}k_{i}^{\mu}$ is the total outflowing four-momentum and the dotted outgoing line symbolizes the background parameters $\{b^{\mu},v^{\mu},\Psi^{\mu}\}$ of \Eqn{backgroundexp}.
We see that only energy is conserved on the worldline. 
The bulk
graviton vertices are generic.  At 4PM order we need the worldline vertices $V_{n|m}$ above for $\{n=1,\ldots,4; m=0,
\ldots,5-n\}$, and the 3-,4-,5-graviton vertices.

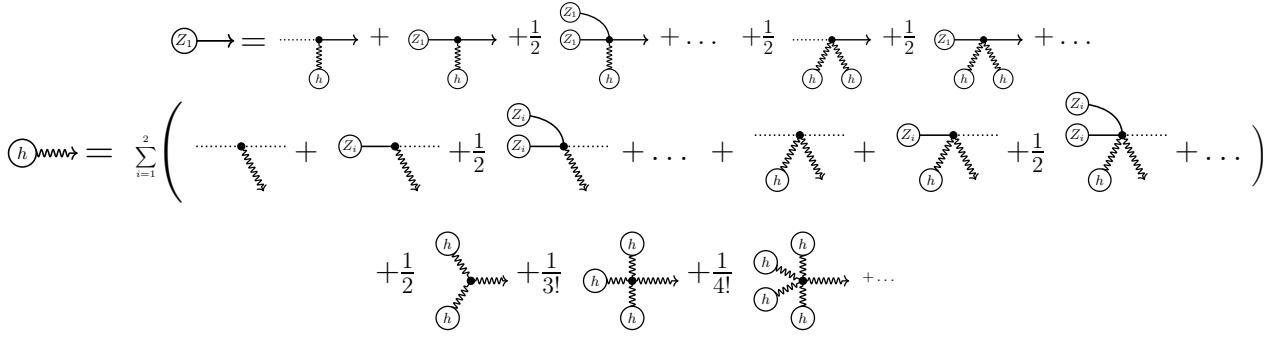
\begin{figure*}[ht!]
 \resizebox{.7\textwidth}{!}{
\resizebox{0.15\textwidth}{!}{\raisebox{20pt}{\begin{tikzpicture}[baseline={(current bounding box.south)}]
    \coordinate (in) at (-1,0);
    \coordinate (x) at (0,0);
    \coordinate (y) at (1,0);
%     \draw[worlddot2] (in) -- (x) ;
    \draw [zUndirected,->] (x) -- (y) ;
        \draw [fill=white, thick] (x) circle (.25) node {$Z_{1}$}; 
  \end{tikzpicture}}
\raisebox{22pt}{\huge =}\, }
\begin{tikzpicture}[baseline={(current bounding box.south)}]
    \coordinate (out) at (0.8,0);
    \coordinate (in1) at (-1,0);
    \coordinate (in2) at (-1,0.5);
    \coordinate (in3) at (-1,0.9);
    \coordinate (x) at (0,0);
    \coordinate (k0) at (0,-1.0) ;
    \coordinate (k1) at (-.9,-1.0) ;
    \coordinate (k2) at (.9,-1.3) ;
    \draw [worlddot2] (in) -- (x) ;
    \draw [zUndirected,->] (x) -- (out1);
%    \draw [zUndirected] (x) to[out=30,in=180] (out3);
    \draw [photon] (k0) -- (x);
%      \draw [photon] (x) -- (k2);
    \draw [fill] (x) circle (.08);
  \draw [fill=white, thick] (k0) circle (.25) node {$h$}; 
    \end{tikzpicture}
\raisebox{33pt}{\, \huge $+$\, }
\begin{tikzpicture}[baseline={(current bounding box.south)}]
    \coordinate (out) at (0.8,0);
    \coordinate (in1) at (-1,0);
    \coordinate (in2) at (-1,0.5);
    \coordinate (in3) at (-1,0.9);
    \coordinate (x) at (0,0);
    \coordinate (k0) at (0,-1.0) ;
    \coordinate (k1) at (-.9,-1.0) ;
    \coordinate (k2) at (.9,-1.3) ;
    \draw [zUndirected] (in) -- (x) ;
    \draw [zUndirected,->] (x) -- (out1);
%    \draw [zUndirected] (x) to[out=30,in=180] (out3);
    \draw [photon] (k0) -- (x);
%      \draw [photon] (x) -- (k2);
    \draw [fill] (x) circle (.08);
 \draw [fill=white, thick] (in) circle (.25) node {$Z_{1}$}; 
  \draw [fill=white, thick] (k0) circle (.25) node {$h$}; 
    \end{tikzpicture}
\raisebox{33pt}{\, \huge $+ \frac{1}{2}$ \,}
\begin{tikzpicture}[baseline={(current bounding box.south)}]
    \coordinate (out) at (0.8,0);
    \coordinate (in) at (-1,0);
    \coordinate (in2) at (-1,0.7);
    \coordinate (in3) at (-1,1.0);
    \coordinate (x) at (0,0);
    \coordinate (k0) at (0,-1.0) ;
    \coordinate (k1) at (-.9,-1.0) ;
    \coordinate (k2) at (.9,-1.3) ;
    \draw [zUndirected] (in) -- (x) ;
    \draw [zUndirected] (in2) to[out=0,in=90] (x);
    \draw [zUndirected,->] (x) -- (out1);
    \draw [photon] (k0) -- (x);
%      \draw [photon] (x) -- (k2);
    \draw [fill] (x) circle (.08);
 \draw [fill=white, thick] (in) circle (.25) node {$Z_{1}$}; 
  \draw [fill=white, thick] (in2) circle (.25) node {$Z_{1}$}; 
  \draw [fill=white, thick] (k0) circle (.25) node {$h$}; 
    \end{tikzpicture}
    \raisebox{33pt}{\, \huge $+ \ldots$ \,}
    \raisebox{33pt}{\, \huge $+ \frac{1}{2}$ \,}
    \begin{tikzpicture}[baseline={(current bounding box.south)}]
    \coordinate (out) at (0.8,0);
    \coordinate (in1) at (-1,0);
    \coordinate (in2) at (-1,0.5);
    \coordinate (in3) at (-1,0.9);
    \coordinate (x) at (0,0);
    \coordinate (k0) at (0,-1.0) ;
    \coordinate (k1) at (-.5,-1.0) ;
    \coordinate (k2) at (.5,-1.0) ;
    \draw [worlddot2] (in) -- (x) ;
    \draw [zUndirected,->] (x) -- (out1);
%    \draw [zUndirected] (x) to[out=30,in=180] (out3);
    \draw [photon] (k1) -- (x);
      \draw [photon] (k2) -- (x);
%      \draw [photon] (x) -- (k2);
    \draw [fill] (x) circle (.08);
  \draw [fill=white, thick] (k1) circle (.25) node {$h$}; 
   \draw [fill=white, thick] (k2) circle (.25) node {$h$}; 
    \end{tikzpicture}
    \raisebox{33pt}{\, \huge $+ \frac{1}{2}$ \,}
    \begin{tikzpicture}[baseline={(current bounding box.south)}]
    \coordinate (out) at (0.8,0);
    \coordinate (in1) at (-1,0);
    \coordinate (in2) at (-1,0.5);
    \coordinate (in3) at (-1,0.9);
    \coordinate (x) at (0,0);
    \coordinate (k0) at (0,-1.0) ;
    \coordinate (k1) at (-.5,-1.0) ;
    \coordinate (k2) at (.5,-1.0) ;
    \draw [zUndirected] (in) -- (x) ;
    \draw [zUndirected,->] (x) -- (out1);
%    \draw [zUndirected] (x) to[out=30,in=180] (out3);
    \draw [photon] (k1) -- (x);
      \draw [photon] (k2) -- (x);
%      \draw [photon] (x) -- (k2);
    \draw [fill] (x) circle (.08);
  \draw [fill=white, thick] (k1) circle (.25) node {$h$}; 
   \draw [fill=white, thick] (k2) circle (.25) node {$h$}; 
    \draw [fill=white, thick] (in) circle (.25) node {$Z_{1}$}; 
    \end{tikzpicture}
    \raisebox{33pt}{\, \huge $+ \ldots$ \,}
}
   \resizebox{0.95\textwidth}{!}{
\resizebox{0.15\textwidth}{!}{\raisebox{20pt}{
\begin{tikzpicture}[baseline={(current bounding box.center)}]
    \coordinate (x) at (0,0);
    \coordinate (y) at (1,0);
%     \draw[worlddot2] (in) -- (x) ;
    \draw [photon,->] (x) -- (y) ;
        \draw [fill=white, thick] (x) circle (.25) node {$h$}; 
  \end{tikzpicture}} \raisebox{15pt}{\huge =}\,  } \raisebox{20pt}{\, 
  \resizebox{0.06\textwidth}{!}{${\displaystyle {\huge \sum_{i=1}^{2}}} {\huge  \Biggl (}$}\, } 
\begin{tikzpicture}[baseline={(current bounding box.south)}]
    \coordinate (out) at (0.8,0);
    \coordinate (in1) at (-1,0);
    \coordinate (in2) at (-1,0.5);
    \coordinate (in3) at (-1,0.9);
    \coordinate (x) at (0,0);
    \coordinate (k0) at (0.5,-1.0) ;
    \coordinate (k1) at (-.9,-1.0) ;
    \coordinate (k2) at (.9,-1.3) ;
    \draw [worlddot2] (in) -- (x) ;
    \draw [worlddot2] (x) -- (out1);
%    \draw [zUndirected] (x) to[out=30,in=180] (out3);
    \draw [photon,->] (x) -- (k0);
%      \draw [photon] (x) -- (k2);
    \draw [fill] (x) circle (.08);
    \end{tikzpicture}
\raisebox{20pt}{\,\huge $+$\, }
\begin{tikzpicture}[baseline={(current bounding box.south)}]
    \coordinate (out) at (0.8,0);
    \coordinate (in1) at (-1,0);
    \coordinate (in2) at (-1,0.5);
    \coordinate (in3) at (-1,0.9);
    \coordinate (x) at (0,0);
    \coordinate (k0) at (0.5,-1.0) ;
    \coordinate (k1) at (-.9,-1.0) ;
    \coordinate (k2) at (.9,-1.3) ;
    \draw [zUndirected] (in) -- (x) ;
    \draw [worlddot2] (x) -- (out1);
%    \draw [zUndirected] (x) to[out=30,in=180] (out3);
    \draw [photon,->] (x) -- (k0);
%      \draw [photon] (x) -- (k2);
    \draw [fill] (x) circle (.08);
 \draw [fill=white, thick] (in) circle (.25) node {$Z_{i}$}; 
%  \draw [fill=white, thick] (k0) circle (.25) node {$h$}; 
    \end{tikzpicture}
\raisebox{20pt}{\,\huge $  + \frac{1}{2}$ \,}
\begin{tikzpicture}[baseline={(current bounding box.south)}]
    \coordinate (out) at (0.8,0);
    \coordinate (in) at (-1,0);
    \coordinate (in2) at (-1,0.7);
    \coordinate (in3) at (-1,1.0);
    \coordinate (x) at (0,0);
    \coordinate (k0) at (0.5,-1.0) ;
    \coordinate (k1) at (-.9,-1.0) ;
    \coordinate (k2) at (.9,-1.3) ;
    \draw [zUndirected] (in) -- (x) ;
    \draw [worlddot2] (x) -- (out1);
    \draw [zUndirected] (in2) to[out=0,in=90] (x);
    \draw [photon,->] (x) -- (k0);
%      \draw [photon] (x) -- (k2);
    \draw [fill] (x) circle (.08);
 \draw [fill=white, thick] (in) circle (.25) node {$Z_{i}$}; 
  \draw [fill=white, thick] (in2) circle (.25) node {$Z_{i}$}; 
%  \draw [fill=white, thick] (k0) circle (.25) node {$h$}; 
    \end{tikzpicture}
    \raisebox{20pt}{\, \huge $+ \ldots$ \,}
    \raisebox{20pt}{\,\huge  $+$ \,}
    \begin{tikzpicture}[baseline={(current bounding box.south)}]
    \coordinate (out) at (0.8,0);
    \coordinate (in1) at (-1,0);
    \coordinate (in2) at (-1,0.5);
    \coordinate (in3) at (-1,0.9);
    \coordinate (x) at (0,0);
    \coordinate (k0) at (0,-1.0) ;
    \coordinate (k1) at (-.5,-1.0) ;
    \coordinate (k2) at (.5,-1.0) ;
    \draw [worlddot2] (in) -- (x) ;
    \draw [worlddot2] (x) -- (out1);
%    \draw [zUndirected] (x) to[out=30,in=180] (out3);
    \draw [photon] (x) -- (k1);
      \draw [photon,->] (x) -- (k2);
%      \draw [photon] (x) -- (k2);
    \draw [fill] (x) circle (.08);
%  \draw [fill=white, thick] (k1) circle (.25) node {$h$}; 
   \draw [fill=white, thick] (k1) circle (.25) node {$h$}; 
    \end{tikzpicture}
    \raisebox{20pt}{\,\huge $+$  \,}
    \begin{tikzpicture}[baseline={(current bounding box.south)}]
    \coordinate (out) at (0.8,0);
    \coordinate (in1) at (-1,0);
    \coordinate (in2) at (-1,0.5);
    \coordinate (in3) at (-1,0.9);
    \coordinate (x) at (0,0);
    \coordinate (k0) at (0,-1.0) ;
    \coordinate (k1) at (-.5,-1.0) ;
    \coordinate (k2) at (.5,-1.0) ;
    \draw [zUndirected] (in) -- (x) ;
    \draw [worlddot2] (x) -- (out1);
%    \draw [zUndirected] (x) to[out=30,in=180] (out3);
    \draw [photon] (x) -- (k1);
      \draw [photon,->] (x) -- (k2);
%      \draw [photon] (x) -- (k2);
    \draw [fill] (x) circle (.08);
%  \draw [fill=white, thick] (k1) circle (.25) node {$h$}; 
   \draw [fill=white, thick] (k1) circle (.25) node {$h$}; 
    \draw [fill=white, thick] (in) circle (.25) node {$Z_{i}$}; 
    \end{tikzpicture}
    \raisebox{20pt}{\,\huge $+ \frac{1}{2}$ \,} 
\begin{tikzpicture}[baseline={(current bounding box.south)}]
    \coordinate (out) at (0.8,0);
    \coordinate (in) at (-1,0);
    \coordinate (in2) at (-1,0.7);
    \coordinate (in3) at (-1,1.0);
    \coordinate (x) at (0,0);
    \coordinate (k0) at (0.5,-1.0) ;
    \coordinate (k1) at (-.5,-1.0) ;
    \coordinate (k2) at (.5,-1.0) ;
    \draw [zUndirected] (in) -- (x) ;
    \draw [worlddot2] (x) -- (out1);
    \draw [zUndirected] (in2) to[out=0,in=90] (x);
    \draw [photon,->] (x) -- (k0);
      \draw [photon] (k1) -- (x);
    \draw [fill] (x) circle (.08);
 \draw [fill=white, thick] (in) circle (.25) node {$Z_{i}$}; 
  \draw [fill=white, thick] (in2) circle (.25) node {$Z_{i}$}; 
  \draw [fill=white, thick] (k1) circle (.25) node {$h$}; 
    \end{tikzpicture}
     \raisebox{20pt}{\,\huge $+ \ldots$  \resizebox{0.02\textwidth}{!}{$ \Biggr )$}\,} }\\[10pt]
  \resizebox{0.4\textwidth}{!}{     \raisebox{30pt}{\,\huge  $+\frac{1}{2}$ \,}
    \begin{tikzpicture}[baseline={(current bounding box.south)}]
    \coordinate (out) at (0.8,0);
    \coordinate (in1) at (-0.5,0.8);
    \coordinate (in2) at (-0.5,-0.8);
    \coordinate (in3) at (-1,1.0);
    \coordinate (x) at (0,0);
    \coordinate (k0) at (0.5,-1.0) ;
    \coordinate (k1) at (-.5,-1.0) ;
    \coordinate (k2) at (.5,-1.0) ;  
    \draw [photon,->] (x) -- (out);
      \draw [photon] (in1) -- (x);
     \draw [photon] (in2) -- (x);
    \draw [fill] (x) circle (.08);
 \draw [fill=white, thick] (in1) circle (.25) node {$h$}; 
  \draw [fill=white, thick] (in2) circle (.25) node {$h$}; 
    \end{tikzpicture}    
    \raisebox{30pt}{\,\huge $+ \frac{1}{3!}$ \,}
    \begin{tikzpicture}[baseline={(current bounding box.south)}]
    \coordinate (out) at (1.0,0);
    \coordinate (in1) at (0,0.8);
    \coordinate (in2) at (0,-0.8);
    \coordinate (in3) at (-0.8,0);
    \coordinate (x) at (0,0);
    \coordinate (k0) at (0.5,-1.0) ;
    \coordinate (k1) at (-.5,-1.0) ;
    \coordinate (k2) at (.5,-1.0) ;  
    \draw [photon,->] (x) -- (out);
      \draw [photon] (in1) -- (x);
     \draw [photon] (in2) -- (x);
          \draw [photon] (in3) -- (x);
    \draw [fill] (x) circle (.08);
 \draw [fill=white, thick] (in1) circle (.25) node {$h$}; 
  \draw [fill=white, thick] (in2) circle (.25) node {$h$}; 
    \draw [fill=white, thick] (in3) circle (.25) node {$h$}; 
    \end{tikzpicture}   
       \raisebox{30pt}{\,\huge $+\frac{1}{4!}$ \,}
    \begin{tikzpicture}[baseline={(current bounding box.south)}]
    \coordinate (out) at (1.0,0);
    \coordinate (in1) at (0,0.8);
    \coordinate (in2) at (0,-0.8);
    \coordinate (in3) at (-0.8,0.4);
     \coordinate (in4) at (-0.8,-0.40);
    \coordinate (x) at (0,0);
    \coordinate (k0) at (0.5,-1.0) ;
    \coordinate (k1) at (-.5,-1.0) ;
    \coordinate (k2) at (.5,-1.0) ;  
    \draw [photon,->] (x) -- (out);
     \draw [photon] (in1) -- (x);
    \draw [photon] (in2) -- (x);
    \draw [photon] (in3) -- (x);
    \draw [photon] (in4) -- (x);
    \draw [fill] (x) circle (.08);
 \draw [fill=white, thick] (in1) circle (.25) node {$h$}; 
  \draw [fill=white, thick] (in2) circle (.25) node {$h$}; 
    \draw [fill=white, thick] (in3) circle (.25) node {$h$}; 
        \draw [fill=white, thick] (in4) circle (.25) node {$h$}; 
    \end{tikzpicture} 
     \raisebox{30pt}{\, $+ \ldots$ \,} 
}
   \caption{\small Berends-Giele type recursion relation to construct $\braket{Z_{i}^{\mu}(\omega)}$ and $\braket{h_{\mu\nu}(k)}$ perturbatively. The causality flow is always from the $Z_{i}$ and $h$ blobs to the outgoing line. They are equivalent to the PM-expanded geodesic and Einstein equations.
  }
  \label{fig:Berends-Giele}
\end{figure*}

\sec{Momentum impulse and spin kick}
The momentum impulse $\Delta p_i^\mu:=[p_i^\mu]^{\tau=+\infty}_{\tau=-\infty}$
and spin kick  $\Delta S_i^{\mu}:=[S_i^\mu]^{\tau=+\infty}_{\tau=-\infty}$ follow from the one-point functions 
\begin{align}	\Delta p_i^\mu&=
	m_i\int_{-\infty}^{\infty}\d\tau\left<\frac{\d^2x_i^\mu(\tau)}{\d\tau^2}\right>
	=-m_i\omega^2\!\left.\braket{z_i^\mu(\omega)}\right|_{\omega=0}\,,\nn \\
	\Delta\psi_i^\mu&=
	\int_{-\infty}^{\infty}\d\tau\left<\frac{\d\psi_i^\mu(\tau)}{\d\tau}\right>
	=-i\omega\!\left.\braket{\psi_i^{\prime\mu}(\omega)}\right|_{\omega=0}\,,
		\label{impulsedef}
\end{align}
where we have Fourier transformed to momentum space.
Both observables are given as the sum of all diagrams at a given PM order with one outgoing $Z_{i}^{\mu}$ line with vanishing energy.
The spin kick is subsequently derived from the kick of the Grassmann variable as in Ref.~\cite{Jakobsen:2022zsx}.

\sec{Integrand generation}
The 4PM impulse and spin-kick integrands are generated recursively via Berends-Giele type relations. The one-point functions for the worldline ``super-fields'' $Z_{i}=\{z_{i},\psi'_{i}\}$ and for the graviton are represented as
\begin{align}
\langle Z_{i}(\omega)\rangle =
  \begin{tikzpicture}[baseline={(current bounding box.center)}]
    \coordinate (in) at (-0.6,0);
    \coordinate (x) at (-.2,0);
    \coordinate (y) at (0.8,0);
    \draw [zUndirected,->] (x) -- (y) node [midway, below] {$\omega$} node [midway, above] {$\rightarrow$};
      \draw [fill=white, thick] (-.4,0) circle (.25) node {$Z_{i}$}; 
  \end{tikzpicture}\, , \quad
  \langle h_{\mu\nu}(k)\rangle =
   \begin{tikzpicture}[baseline={(current bounding box.center)}]
    \coordinate (in) at (-0.6,0);
    \coordinate (x) at (-.2,0);
    \coordinate (y) at (0.8,0);
    \draw [photon,->] (x) -- (y) node [midway, above] {$\rightarrow$} node [midway, below] {$k$};
   \draw [fill=white,thick] (-.4,0) circle (.25) node {$h$}; 
  \end{tikzpicture}\,.
\end{align}
Their recursive definitions follow from the Schwinger-Dyson equations and are depicted in \Fig{fig:Berends-Giele}.
Spelling this out systematically to order $G^{4}$ allows for an algorithmic construction
of the integrand:
in our case, we efficiently inserted  Feynman rules into the generated trees using {\tt FORM} \cite{Ruijl:2017dtg}.
There are 201 graphs contributing to the 4PM impulse in the non-spinning case,
529 with spin and 253 contributing to the 4PM spin kick.

\sec{Reduction to scalar integrals}
A generic 4PM diagram after performing the worldline energy integrals
via the $\dd$-functions in \Eqn{eq:WLvertexrule} takes the form
\begin{align}\label{intfamily}
&\int_{q} e^{-i q\cdot b}  \dd(q\cdot v_{1}) \dd(q\cdot v_{2})  \times\\ &\qquad
\int_{\ell_{1},\ell_{2},\ell_{3}} \!\frac{\text{num$[\ell_{i}]$}}{D_{1}\cdots D_{12}}
  \dd(\ell_{1}\cdot v_{i_{1}}) \dd(\ell_{2}\cdot v_{i_{2}}) \dd(\ell_{3}\cdot v_{i_{3}})\,, \nn
\end{align}
where the $D_{i}$ are either linear or massless  propagators depending on the loop momenta 
$\ell_{i}$, velocities $v_{i}$ and momentum transfer $q$. 
The numerators $\text{num$[\ell_{i}]$}$ are polynomial in loop momenta.
Tensor reduction of $\text{num}[\ell_{i}]$ to scalar integrals is performed by expanding
the loop momenta on a basis dual to $v_i^\mu$ and $q^\mu$, as demonstrated in the 3PM case \cite{Jakobsen:2022fcj}.
The only dimensionful quantity in the 3-loop $\ell_{i}$ integral is the momentum transfer $q^{\mu}$.
Hence, $|q|=\sqrt{-q^{2}}$ may be scaled out,
and the remaining 3-loop integrals depend only on the Lorentz factor $\gamma$. 
\begin{figure*}[ht!]
  \resizebox{0.95\textwidth}{!}{
    \begin{tikzpicture}[baseline={([yshift=-1ex]current bounding box.south)},scale=.7]
    \coordinate (inA) at (0.4,.7);
    \coordinate (outA) at (3.6,.7);
    \coordinate (inB) at (0.4,-.7);
    \coordinate (outB) at (3.6,-.7);
    \coordinate (xA) at (1,.7);
     \coordinate (yA) at (2,.7);
     \coordinate (zA) at (3,.7);
     \coordinate (xB) at (1,-.7);
    \coordinate (yB) at (2,-.7);
    \coordinate (zB) at (3,-.7);
    \coordinate (xM) at (1,0);
    \coordinate (yM) at (2,0);
    \coordinate (zM) at (3,-0);

     %%%
    \draw [fill] (xA) circle (.08);
%    \draw [fill] (yA) circle (.08);
    \draw [fill] (zA) circle (.08);
    \draw [fill] (xB) circle (.08);
    \draw [fill] (yB) circle (.08);
    \draw [fill] (zB) circle (.08);
    \draw [fill] (xM) circle (.08);
    \draw [fill] (yM) circle (.08);
    \draw [fill] (zM) circle (.08);
    %%%
    \draw [dotted] (inA) -- (outA);
    \draw [dotted] (inB) -- (outB);
    % \draw [photon] (zA) -- (outA);
    \draw [draw=none] (xA) to[out=40,in=140] (zA);
    %%%
%    \draw [zUndirected] (xA) -- (yzA);
%    \draw [zUndirected] (yB) -- ();
    \draw [photon] (xA) -- (xM);
    \draw [photon] (xM) -- (xB);
    \draw [photon] (zA) -- (zM);
    \draw [photon] (zM) -- (zB);
    \draw [photon] (yB) -- (yM);
    \draw [photon] (xM) -- (yM);
    \draw [photon] (yM) -- (zM);
  \end{tikzpicture}
    \begin{tikzpicture}[baseline={([yshift=-1ex]current bounding box.south)},scale=.7]
    \coordinate (inA) at (0.4,.7);
    \coordinate (outA) at (3.6,.7);
    \coordinate (inB) at (0.4,-.7);
    \coordinate (outB) at (3.6,-.7);
    \coordinate (xA) at (1,.7);
     \coordinate (yA) at (2,.7);
     \coordinate (zA) at (3,.7);
     \coordinate (xB) at (1,-.7);
    \coordinate (yB) at (2,-.7);
    \coordinate (zB) at (3,-.7);
    \coordinate (xM) at (1,0);
    \coordinate (yM) at (2,0);
    \coordinate (zM) at (3,-0);

     %%%
    \draw [fill] (xA) circle (.08);
    \draw [fill] (yA) circle (.08);
    \draw [fill] (zA) circle (.08);
    \draw [fill] (xB) circle (.08);
    \draw [fill] (yB) circle (.08);
    \draw [fill] (zB) circle (.08);
    \draw [fill] (xM) circle (.08);
    \draw [fill] (yM) circle (.08);
    \draw [fill] (zM) circle (.08);
    %%%
    \draw [dotted] (inA) -- (outA);
    \draw [dotted] (inB) -- (outB);
    % \draw [photon] (zA) -- (outA);
    \draw [draw=none] (xA) to[out=40,in=140] (zA);
    %%%
%    \draw [zUndirected] (xA) -- (yzA);
%    \draw [zUndirected] (yB) -- ();
    \draw [photon] (xA) -- (xM);
    \draw [photon] (xM) -- (xB);
    \draw [photon] (zA) -- (zM);
    \draw [photon] (zM) -- (zB);
    \draw [photon] (yB) -- (yM);
    \draw [photon] (xM) -- (yM);
    \draw [photon] (yM) -- (yA);
    \draw [zUndirected] (yA) -- (zA);
  \end{tikzpicture}
 \begin{tikzpicture}[baseline={([yshift=-1ex]current bounding box.south)},scale=.7]
    \coordinate (inA) at (0.4,.7);
    \coordinate (outA) at (4.6,.7);
    \coordinate (inB) at (0.4,-.7);
    \coordinate (outB) at (4.6,-.7);
    \coordinate (xA) at (1,.7);
     \coordinate (yA) at (2,.7);
     \coordinate (zA) at (3,.7);
     \coordinate (wA) at (4,.7);
    \coordinate (xB) at (1,-.7);
    \coordinate (yB) at (2,-.7);
    \coordinate (zB) at (3,-.7);
    \coordinate (wB) at (4,-.7);
    \coordinate (xM) at (1,0);
    \coordinate (yM) at (2,0);
    \coordinate (zM) at (3,-0);

     %%%
    \draw [fill] (xA) circle (.08);
    \draw [fill] (yA) circle (.08);
    \draw [fill] (zA) circle (.08);
    \draw [fill] (xB) circle (.08);
    \draw [fill] (yB) circle (.08);
    \draw [fill] (zB) circle (.08);
    \draw [fill] (wB) circle (.08);
    \draw [fill] (wA) circle (.08);
    %%%
    \draw [dotted] (inA) -- (outA);
    \draw [dotted] (inB) -- (outB);
    % \draw [photon] (zA) -- (outA);
    \draw [draw=none] (xA) to[out=40,in=140] (zA);
    %%%
%    \draw [zUndirected] (xA) -- (yzA);
%    \draw [zUndirected] (yB) -- ();
    \draw [photon] (xA) -- (xB);
    \draw [photon] (yA) -- (yB);
    \draw [photon] (zA) -- (zB);
    \draw [photon] (wA) -- (wB);
    \draw [zUndirected] (xA) -- (yA);
    \draw [zUndirected] (zA) -- (wA);
    \draw [zUndirected] (yB) -- (zB);
  \end{tikzpicture}
  \begin{tikzpicture}[baseline={([yshift=-1ex]current bounding box.south)},scale=.7]
    \coordinate (inA) at (0.4,.7);
    \coordinate (outA) at (4.6,.7);
    \coordinate (inB) at (0.4,-.7);
    \coordinate (outB) at (4.6,-.7);
    \coordinate (xA) at (1,.7);
     \coordinate (yA) at (2,.7);
     \coordinate (zA) at (3,.7);
     \coordinate (wA) at (4,.7);
    \coordinate (xB) at (1,-.7);
    \coordinate (yB) at (2,-.7);
    \coordinate (zB) at (3,-.7);
    \coordinate (wB) at (4,-.7);
    \coordinate (xM) at (1,0);
    \coordinate (yM) at (2,0);
    \coordinate (zM) at (3,-0);

     %%%
    \draw [fill] (xA) circle (.08);
    \draw [fill] (yA) circle (.08);
    \draw [fill] (zA) circle (.08);
    \draw [fill] (xB) circle (.08);
    \draw [fill] (yB) circle (.08);
    \draw [fill] (zB) circle (.08);
    \draw [fill] (wB) circle (.08);
    \draw [fill] (wA) circle (.08);
    %%%
    \draw [dotted] (inA) -- (outA);
    \draw [dotted] (inB) -- (outB);
    % \draw [photon] (zA) -- (outA);
    \draw [draw=none] (xA) to[out=40,in=140] (zA);
    %%%
%    \draw [zUndirected] (xA) -- (yzA);
%    \draw [zUndirected] (yB) -- ();
    \draw [photon] (xA) -- (xB);
    \draw [photon] (yA) -- (yB);
    \draw [photon] (zA) -- (zB);
    \draw [photon] (wA) -- (wB);
    \draw [zUndirected] (xB) -- (yB);
    \draw [zUndirected] (zA) -- (wA);
    \draw [zUndirected] (yA) -- (zA);
  \end{tikzpicture}
 \begin{tikzpicture}[baseline={([yshift=-1ex]current bounding box.south)},scale=.7]
    \coordinate (inA) at (0.4,.7);
    \coordinate (outA) at (4.6,.7);
    \coordinate (inB) at (0.4,-.7);
    \coordinate (outB) at (4.6,-.7);
    \coordinate (xA) at (1,.7);
     \coordinate (yA) at (2.5,.7);
     \coordinate (zA) at (4,.7);
     \coordinate (xB) at (1,-.7);
    \coordinate (yB) at (2.5,-.7);
    \coordinate (zB) at (4,-.7);
    \coordinate (xM) at (1,0);
    \coordinate (yM) at (2,0);
    \coordinate (zM) at (3,-0);

     %%%
    \draw [fill] (xA) circle (.08);
    \draw [fill] (yA) circle (.08);
    \draw [fill] (zA) circle (.08);
    \draw [fill] (xB) circle (.08);
    \draw [fill] (yB) circle (.08);
    \draw [fill] (zB) circle (.08);
    %%%
    \draw [dotted] (inA) -- (outA);
    \draw [dotted] (inB) -- (outB);
    % \draw [photon] (zA) -- (outA);
    \draw [draw=none] (xA) to[out=40,in=140] (zA);
    %%%
%    \draw [zUndirected] (xA) -- (yzA);
%    \draw [zUndirected] (yB) -- ();
    \draw [photon] (xA) -- (xB);
    \draw [photon] (yA) -- (yB);
    \draw [photon] (zA) -- (zB);
    \draw [photon] (xB) to[out=80,in=100] (yB);
    \draw [zUndirected] (yA) -- (zA);
  \end{tikzpicture}
 \begin{tikzpicture}[baseline={([yshift=-1ex]current bounding box.south)},scale=.7]
    \coordinate (inA) at (0.4,.7);
    \coordinate (outA) at (4.6,.7);
    \coordinate (inB) at (0.4,-.7);
    \coordinate (outB) at (4.6,-.7);
    \coordinate (xA) at (1,.7);
     \coordinate (yA) at (2.5,.7);
     \coordinate (zA) at (4,.7);
     \coordinate (xB) at (1,-.7);
    \coordinate (yB) at (2.5,-.7);
    \coordinate (zB) at (4,-.7);
    \coordinate (xM) at (1,0);
    \coordinate (yM) at (2.5,0);
    \coordinate (zM) at (4,-0);

     %%%
    \draw [fill] (xA) circle (.08);
    \draw [fill] (yA) circle (.08);
    \draw [fill] (zA) circle (.08);
    \draw [fill] (xB) circle (.08);
    \draw [fill] (yB) circle (.08);
    \draw [fill] (zB) circle (.08);
    %%%
    \draw [dotted] (inA) -- (outA);
    \draw [dotted] (inB) -- (outB);
    % \draw [photon] (zA) -- (outA);
    \draw [draw=none] (xA) to[out=40,in=140] (zA);
    %%%
%    \draw [zUndirected] (xA) -- (yzA);
%    \draw [zUndirected] (yB) -- ();
    \draw [photon] (xA) -- (xB);
    \draw [photon] (yA) -- (yB);
    \draw [photon] (zA) -- (zB);
    \draw [photon] (xA) to[out=-80,in=-100] (yA);
    \draw [zUndirected] (yA) -- (zA);
  \end{tikzpicture}
  \begin{tikzpicture}[baseline={([yshift=-1ex]current bounding box.south)},scale=.7]
    \coordinate (inA) at (0.4,.7);
    \coordinate (outA) at (4.6,.7);
    \coordinate (inB) at (0.4,-.7);
    \coordinate (outB) at (4.6,-.7);
    \coordinate (xA) at (1,.7);
    \coordinate (xyAB) at (1.5,0) ;
    \coordinate (yB) at (2,-.7);
     \coordinate (yA) at (2,.7);
     \coordinate (zA) at (3,.7);
     \coordinate (wA) at (4,.7);
    \coordinate (xB) at (1,-.7);
    \coordinate (zB) at (3,-.7);
    \coordinate (wB) at (4,-.7);
    \coordinate (xM) at (1,0);
    \coordinate (yM) at (2,0);
    \coordinate (zM) at (3,-0);

     %%%
    \draw [fill] (xA) circle (.08);
    \draw [fill] (yA) circle (.08);
    \draw [fill] (zA) circle (.08);
    \draw [fill] (xB) circle (.08);
    \draw [fill] (yB) circle (.08);
    \draw [fill] (zB) circle (.08);
    \draw [fill] (wB) circle (.08);
    \draw [fill] (wA) circle (.08);
    %%%
    \draw [dotted] (inA) -- (outA);
    \draw [dotted] (inB) -- (outB);
    % \draw [photon] (zA) -- (outA);
    \draw [draw=none] (xA) to[out=40,in=140] (zA);
    %%%
%    \draw [zUndirected] (xA) -- (yzA);
%    \draw [zUndirected] (yB) -- ();
    \draw [photon] (xA) -- (yB);
    \filldraw[white] (xyAB) circle (6pt) ;
%    \node [cicle,color=white,line width=4] () at (middle) ;
    \draw [photon] (yA) -- (xB);
    \draw [photon] (zA) -- (zB);
    \draw [photon] (wA) -- (wB);
    \draw [zUndirected] (zB) -- (yB);
    \draw [zUndirected] (zA) -- (wA);
    \draw [zUndirected] (yA) -- (zA);
  \end{tikzpicture}
}
   \caption{\small
  	Examples of comparable-mass graphs  
	with mass dependence $m_{1}^{2} m_{2}^{3}$
	contributing to the 4PM calculation. One should attach an outgoing worldline to \emph{any} worldline node and apply the resulting causality flow. The corresponding scalar integrals
	feature as top sectors in the differential equations: 
	All graphs can be described by the J family~\eqref{eq:jFamily}, except for the last 
	graph belonging to the I family~\eqref{eq:general}. The first two graphs give rise to the elliptic functions in the final result. The second-to-last graph is non-zero only in the (PR+RP) region and 
	therefore does not contribute to the conservative results in this paper.   }
  \label{fig:equalmassgraphs}
\end{figure*}
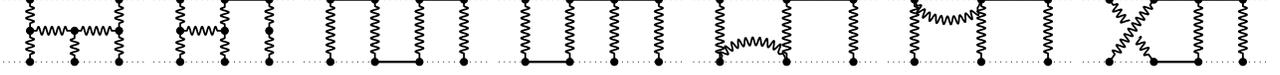

The specific choice of three $\dd(\ell_k\cdot v_{i_k})$ functions in \Eqn{intfamily}
follows the mass dependence of a given diagram,
which scales as $m_{1}m_{2}m_{i_{1}}m_{i_{2}}m_{i_{3}}$.
Diagrams are thereby grouped into two categories: test-body
contributions with mass dependence $m_1^4m_2$ or $m_1m_2^4$
%(see fig.~\ref{fig:testbodygraphs})
and comparable-mass contributions $m_1^3m_2^2$, $m_1^2m_2^3$ ---
see \Fig{fig:equalmassgraphs}.
For the conservative impulse we can easily reconstruct the $m_1m_2^4$ and $m_1^2m_2^3$
components using $\Delta p_{1,{\rm cons}}^\mu=-\Delta p_{2,{\rm cons}}^\mu$,
the impulse on the second body being given simply by relabeling the two worldlines.
When computing $\Delta\psi_{1,{\rm cons}}^\mu$ no similar relation exists;
however, the integrals in opposing mass sectors are also related by a trivial relabeling.
 
\sec{Integral families and reduction to masters}
There are three integral families that need to be reduced to master integrals. 
The first 4PM family is $(i=1,2)$
\begin{subequations}
  \label{eq:general}
  \begin{align}
    I^{[i](\sigma_1,\sigma_2,\sigma_3)}_{n_1,n_2,...,n_{12}}
    =
    \int_{\ell_1,\ell_2,\ell_3}\!\!\!\!\!\!
    \frac{
      \dd(\ell_1\cdot v_{i})
      \dd(\ell_2\cdot v_{1})
      \dd(\ell_3\cdot v_{1})
    }{
      D_1^{n_1}
      D_2^{n_2}
      ...
      D_{12}^{n_{12}}
    }
  \end{align}
  with the propagators ($j=1,2,3$ and $k=1,2$):
  \begin{align}
    &D_{1}
    =
    \ell_1\cdot v_{\bar\imath} +\sigma_1i0^+
    \,,\,\,
    D_{1+k}
    =
    \ell_{1+k}\cdot v_2 +\sigma_{1+k}i0^+\,,\nn
    \\
    &D_{4}
    =
    (\ell_1+\ell_2+\ell_3+q)^2
    \,,\,\,
    D_{5}
    =
    (\ell_1+\ell_2+q)^2
    \,,
    \\
    &D_{5+k}\!
    =
    (\ell_k+\ell_3)^2
    \,,\,\,
    D_{7+j}
    =
    \ell_{j}^2
    \,,\,\,
    D_{10+k}\!
    =
    (\ell_{k}+q)^2\, ,\nn
  \end{align}
\end{subequations}
and $\bar1=2$, $\bar2=1$.
The $I^{[1]}$ and $I^{[2]}$ families contribute to the test-body
and comparable-mass regimes respectively.
The other 4PM family is given by
\begin{subequations}\label{eq:jFamily}
  \begin{align}
    J^{(\sigma_1,\sigma_2,\sigma_3)}_{n_1,n_2,\ldots,n_{12}}
    :=
    \int_{\ell_1,\ell_2,\ell_3}\!\!\!\!
    \frac{
      \dd(\ell_1\cdot v_{1})
      \dd(\ell_2\cdot v_{1})
      \dd(\ell_3\cdot v_{2})
    }{
      D_1^{n_1}
      D_2^{n_2}
      ...
      D_{12}^{n_{12}}
    }
  \end{align}
  with ($j=1,2,3$, $k=1,2$)
  \begin{align}
    &
    D_k
    =
    \ell_{k}\cdot v_{2} +\sigma_{k}i0^+
    \,,\,\,
    D_3
    =
    \ell_{3}\cdot v_{1} +\sigma_{3}i0^+\,,
    \nn\\
    &
    D_{3+k}
    =
    (\ell_k-\ell_3)^2
    \,,\,\,
    D_{6}
    =
    (\ell_1-\ell_2)^2
    \, ,
   \\
   &
   D_{6+j}
   =\ell_j^2
   \,,\,\,
   D_{9+j}
   =(\ell_j+q)^2
   \,.\nn
  \end{align}
\end{subequations}
Each family splits into two branches: even ($b$-type) or odd ($v$-type)
in the number of worldline propagators.
In the non-spinning impulse, these integrals multiply terms proportional to $b^\mu$, $v_i^\mu$ respectively
\eqref{eq:bvtype}. 
Using integration-by-parts (IBP) relations \cite{Lee:2013mka,Smirnov:2019qkx,Maierhofer:2017gsa,Klappert:2020nbg}
we reduce the families to   
23 master integrals for the I-$b$ and I-$v$ types each as well as 64 of J-$b$ type
and 66 of J-$v$ type. 
The complete spinning impulse computation (including dissipation) results in approximately
10$^5$ integrals for reduction to scalar masters.

\sec{Differential equations} To solve for the master integrals we employ the method of canonical differential equations (DEs) \cite{Henn:2013pwa}. Each master integral family 
is grouped into  a vector $\vec{I}$ ordered  according to the number of active propagators. The DE in $x=\gamma - \sqrt{\gamma^{2}-1}$ reads
$ 
\d\vec{I}/\d x = M(\epsilon, x)\, \vec{I}
$
with a lower-block triangular matrix $M(\epsilon, x)$. Finding a transformation matrix $T$ 
that brings us to a canonical basis with an $\epsilon$ factorized DE 
$
\d\vec{\tilde I}/\d x = \epsilon A(x)\, \vec{\tilde I}
$
is a highly involved procedure in which we employ the packages 
\cite{Dlapa:2022wdu,Meyer:2017joq,Prausa:2017ltv,Lee:2021}.
The resulting symbol alphabet is $\{x,1+x,1-x,1+x^{2}\}$, and
we encounter elliptic integrals in the J-$b$ family \cite{Dlapa:2022wdu,Dlapa:2023hsl}.

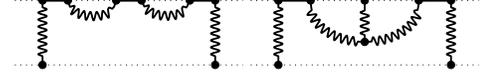
\begin{figure}[ht!]
  \resizebox{0.35\textwidth}{!}{\begin{tikzpicture}[baseline={([yshift=-1ex]current bounding box.south)},scale=.7]
  \coordinate (inA) at (0.4,.7);
  \coordinate (outA) at (5.35,.7);
  \coordinate (inB) at (0.4,-.7);
  \coordinate (outB) at (5.35,-.7);
  \coordinate (xA) at (1,.7);
  \coordinate (xxA) at (1.55,.7) ;
  \coordinate (yA) at (2.60,.7);
  \coordinate (yyA) at (3.15,.7) ;
  \coordinate (zA) at (4.20,.7);
  \coordinate (zzA) at (4.75,.7) ;
  \coordinate (xB) at (1,-.7);
  \coordinate (yB) at (2.5,-.7);
  \coordinate (zB) at (4,-.7);
  \coordinate (zzB) at (4.75,-.7);
  \coordinate (xM) at (1,0);
  \coordinate (yM) at (2.5,0);
  \coordinate (zM) at (4,-0);

  %%%
  \draw [fill] (xA) circle (.08);
  \draw [fill] (xxA) circle (.08);
  \draw [fill] (yA) circle (.08);
  \draw [fill] (yyA) circle (.08);
  \draw [fill] (zA) circle (.08);
  \draw [fill] (zzA) circle (.08);
  \draw [fill] (xB) circle (.08);
  \draw [fill] (zzB) circle (.08);
  %%%
  \draw [dotted] (inA) -- (outA);
  \draw [dotted] (inB) -- (outB);
  % \draw [photon] (zA) -- (outA);
  \draw [draw=none] (xA) to[out=40,in=140] (zA);
  %%%
  %    \draw [zUndirected] (xA) -- (yzA);
  %    \draw [zUndirected] (yB) -- ();
  \draw [photon] (xA) -- (xB);
  \draw [photon] (zzA) -- (zzB);
  \draw [photon] (xxA) to[out=-90,in=-90] (yA);
  \draw [photon] (yyA) to[out=-80,in=-100] (zA);
  \draw [zUndirected] (xA) -- (xxA);
  \draw [zUndirected] (yA) -- (yyA);
  \draw [zUndirected] (zA) -- (zzA);
\end{tikzpicture}
\begin{tikzpicture}[baseline={([yshift=-1ex]current bounding box.south)},scale=.7]
  \coordinate (inA) at (0.4,.7);
  \coordinate (outA) at (5.35,.7);
  \coordinate (inB) at (0.4,-.7);
  \coordinate (outB) at (5.35,-.7);
  \coordinate (xA) at (1,.7);
  \coordinate (xxA) at (1.70,.7) ;
  \coordinate (yA) at (2.875,.7) ;
  \coordinate (zA) at (4.05,.7);
  \coordinate (zzA) at (4.75,.7) ;
  \coordinate (xB) at (1,-.7);
  \coordinate (yB) at (2.5,-.7);
  \coordinate (zB) at (4,-.7);
  \coordinate (zzB) at (4.75,-.7);
  \coordinate (xM) at (1,0);
  \coordinate (yM) at (2.875,-.2);
  \coordinate (zM) at (4,-0);

  %%%
  \draw [fill] (xA) circle (.08);
  \draw [fill] (xxA) circle (.08);
  \draw [fill] (yA) circle (.08);
  \draw [fill] (zA) circle (.08);
  \draw [fill] (zzA) circle (.08);
  \draw [fill] (xB) circle (.08);
  \draw [fill] (zzB) circle (.08);
  \draw [fill] (yM) circle (.08);
  %%%
  \draw [dotted] (inA) -- (outA);
  \draw [dotted] (inB) -- (outB);
  % \draw [photon] (zA) -- (outA);
  \draw [draw=none] (xA) to[out=40,in=140] (zA);
  %%%
  %    \draw [zUndirected] (xA) -- (yzA);
  %    \draw [zUndirected] (yB) -- ();
  \draw [photon] (xA) -- (xB);
  \draw [photon] (zzA) -- (zzB);
  \draw [photon] (yM) to[out=-180,in=-90] (xxA);
  %\draw [photon] (xxA) to[out=-90,in=-180] (yM);
  \draw [photon] (yM) to[out=0,in=-90] (zA);
  \draw [photon] (yM) -- (yA);
  \draw [zUndirected] (xA) -- (xxA);
  \draw [zUndirected] (zA) -- (zzA);
\end{tikzpicture}
}
  \caption{\small
    Two examples of graphs contributing in the (RR) region but not the (PP) region.
   }
  \label{fig:rrGraphs}
\end{figure}

\sec{Fixing boundary conditions } Boundary conditions on the master integrals are determined in the 
static limit ($\gamma\to 1$, $v\to 0$) using the \emph{method of regions} \cite{Smirnov:2012gma,Bern:2021yeh,Herrmann:2021tct,Becher:2014oda} to expand
the integrand in $v$.
Regions in the static limit are characterized by different scalings of the bulk graviton loop momenta
with potential (P) and radiative (R) modes defined by relative scalings of their spacial and timelike components:
\be
	  \ell^\text{P}_i=(\ell_i^0,\Bell_i)\sim(v,1)\,,  \qquad
	  \ell_i^\text{R}=(\ell_i^0,\Bell_i)\sim(v,v)\, .
\ee
Only gravitons which may go on-shell can be radiative and there are at most two of these defining the three regions:
(PP), (RR) and (PR+RP).
The regions (PP) and (PR+RP) are purely conservative and dissipative respectively,
while the (RR) region carries both kinds of effects.
In the (PP) region the integrals reduce to test-body integrals described by the $I^{[1]}$ family~\eqref{eq:general};
in the (RR) region they reduce to double-mushroom integrals like the first graph of Fig.~\ref{fig:rrGraphs}.
Either way, the boundary integrals are independent of $\gamma$ and thus functions only of $D=4-2\eps$.
In our conservative observables we include only the (PP) and (RR) regions as the (PR+RP) region generates terms odd in $v$.

Reaching 4PM order introduces the physical phenomenon of tail effects~\cite{Bern:2021yeh,Dlapa:2021vgp}.
In the 4PM contributions to an observable $X$ (impulse, spin kick or scattering angle)
poles in $\eps=\frac{4-D}{2}$ appear in the (PP) and conservative (RR) contributions:
\begin{align}
  X^{(\rm PP)}
  =
  \frac{P(\gamma)}{2\eps} 
  +\ldots,
  \,\,
  X^{(\rm RR)}_{\rm cons}
  =
  -v^{-4\eps}\frac{P(\gamma)}{2\eps} 
  +\ldots,
\end{align}
higher-order terms being finite as $\eps\to0$.
Non-analytic dependence on $v^{-4\eps}$ in the (RR) region is a direct consequence
of the velocity scaling of the two radiative gravitons.
The cancelation of these poles when assembling $X_{\rm cons}$ introduces logarithmic velocity dependence:
\begin{align}
  X_{\rm cons}
  =
  X^{\rm (PP)}
  +
  X^{\rm (RR)}_{\rm cons}
  =
  P(\gamma) \log\frac{\gamma-1}{2}
  +
  \ldots\,,
\end{align}
the dots indicating terms that are rational in $\sqrt{\gamma^2-1}$ in the static limit.

\sec{Results}
We begin with $\Delta p_{i,\rm cons}^{(4)\mu}$, the $G^4$ component of the impulse $\Delta p_{i,\rm cons}^{\mu}$.
It may be decomposed as
\begin{align}
  &\Delta p_{\text{cons},1}^{(4)\mu} = 
  \frac{m_{1}^2m_{2}^2}{|b|^{4}}
  \sum_{l,\sigma=b,v }
  \rho^{(\sigma)\mu}_{l}
  \bigg[
    \Big(
    \frac{m_{2}^{2}}{m_1}
    c_{l}^{(\sigma)}(\gamma)
    +
    \frac{m_{1}^{2}}{m_2}
    \bar{c}_{l}^{(\sigma)}(\gamma)
    \Big) 
    \nn \\
    &\qquad+
    \sum_{\alpha }
    F^{(\sigma)}_{\alpha}(\gamma)
    \Big(
    m_2
    d_{ \alpha, l}^{(\sigma)}(\gamma) 
    +
    m_1
    \bar{d}_{\alpha,l}^{(\sigma)}(\gamma)
    \Big)
    \bigg]\,,
\end{align}
where the basis vectors and spin structures $\rho^{(b,v)\mu}_{l}$ are
 \begin{align}\label{eq:bvtype}
   \begin{split}
     \rho^{(b)\mu}_{l}
     &=
     \bigg\{
     \hat b^{\mu} , \frac{a_{i}\cdot \hat L}{|b|}\, \hat b^{\mu},
     \frac{a_{i}\cdot \hat b}{|b|}\, \hat L^{\mu}
     \bigg\}
     \ ,
     \\
     \rho^{(v)\mu}_{l}
     &=
     \bigg \{
     v_{j}^{\mu} ,
     \frac{a_{i}\cdot \hat L}{|b|}\,  v_{j}^{\mu},
     \frac{a_{i}\cdot { v}_{\bar\imath}}{|b|}\, \hat L^{\mu}
     \bigg \} \, .
   \end{split}
 \end{align}
There are five and eight elements in $\rho^{(b)\mu}_{l}$ and $\rho^{(v)\mu}_{l}$ respectively,
and the normalized angular momentum
$\hat L^\mu=\eps^{\,\,\,\,\,\,\,\,\,\,\mu}_{\nu\rho\sigma}v_{1}^\nu v_{2}^{\rho}\hat b^\sigma/\gamma v$.
The $ c_{l}^{(\sigma)}(\gamma)$ and $ d_{\alpha, l}^{(\sigma)}(\gamma) $ 
and their barred counterparts are rational functions 
(up to integer powers of $\sqrt{\gamma^{2}-1}$).
All non-trivial dependence on $\gamma$ is contained in the
 16 functions $F^{(b)}_{\alpha}(\gamma)$ with $\gamma_\pm=\gamma\pm1$:
 \begin{align}\label{eq:functions}
&F^{(b)}_{\alpha}(\gamma)=\{1,\text{arccosh}[\gamma], \log [\gamma], \log \left[\frac{\gamma_{\pm}}{2} 
 \right],
 \nn \\ &\quad
 \text{arccosh}^{2}[\gamma], 
 \text{arccosh}[\gamma] \log \left[\frac{\gamma_{\pm}}{2}
 \right ],
  \log \left[\frac{\gamma_{+}}{2} \right ]\log \left[\frac{\gamma_{-}}{2} \right ], \nn \\& \quad
   \log^{2} \left[\frac{\gamma_{+}}{2}\right ],
    \text{Li}_{2}\left [\pm\frac{\gamma_{-}}{\gamma_{+}}\right], \,
 \text{Li}_{2}\left [\sqrt{\frac{\gamma_{-}}{\gamma_{+}}}\right], \\ &\quad
   \eK^2\left [ \frac{\gamma_{-}}{\gamma_{+}} \right ],
   \eE^2\left [ \frac{\gamma_{-}}{\gamma_{+}} \right ],
   \eK\left [ \frac{\gamma_{-}}{\gamma_{+}} \right ]
   \eE\left [ \frac{\gamma_{-}}{\gamma_{+}} \right ]
   \}
   \ ,
   \nn
 \end{align}
and  the much simpler set $F_{\alpha}^{(v)}=\{ 1,\text{arccosh}[\gamma] \} $.
The first line of Eq.~\eqref{eq:functions} includes transcendental weight-1 functions,
the second and third lines weight-2 functions and the final line quadratic combinations
of elliptic functions of the first and second kind.
The barred coefficients $\bar c_l^{(\sigma)}$ and $\bar d_l^{\sigma}$ may be obtained from the unbarred ones by relabeling using  
$\Delta p_{\text{cons},1}^{(4)\mu} = -\Delta p_{\text{cons},2}^{(4)\mu}$.
 
The $G^4$ component of the spin kick $\Delta S_{i,\rm cons}^{(4)\mu}$ admits a similar decomposition,
involving the same functions $F_{\alpha}^{(b,v)}$ but a different set of basis vectors and spin structures:
\begin{align}%\label{eq:bvtype}
\begin{split}
     \tilde{\rho}^{(b)\mu}
     &=
     \bigg\{
     \frac{a_1\cdot v_2}{|b|}
     \hat b^{\mu}
     ,
     \frac{a_1 \cdot \hat{b}}{|b|}
     v_j^\mu
     \bigg\}\,, \\
     \tilde{\rho}^{(v)\mu}
     &=
     \bigg \{
     \frac{a_1 \cdot \hat{b}}{|b|}
     \hat{b}^\mu
     ,
     \frac{a_1 \cdot v_2}{|b|}
     v_j^\mu
     \bigg \} \,,
     \end{split}
 \end{align}
and takes the schematic form
\begin{align}
  \Delta S_{\text{cons},1}^{(4)\mu} &= 
  \frac{m_{1}^2m_{2}^2}{|b|^{3}}
  \sum_{l,\sigma }
  \tilde{\rho}^{(\sigma)\mu}_{l}
  \bigg[
    \Big(
    \frac{m_{2}^{2}}{m_1}
    e_{l}^{(\sigma)}(\gamma)
    +
    \frac{m_{1}^{2}}{m_2}
    \bar{e}_{l}^{(\sigma)}(\gamma)
    \Big) \nn \\
    +&
    \sum_{\alpha }
    F^{( \sigma)}_{\alpha}(\gamma)
    \Big(
    m_2
    f_{\alpha,l}^{(\sigma)}(\gamma)
    +
    m_1
    \bar{f}_{\alpha, l}^{(\sigma)}(\gamma)
    \Big)
    \bigg]\, .
\end{align}
Here $e_{l}^{(\sigma)}(\gamma)$ and $f_{\alpha, l}^{(\sigma)}(\gamma) $ 
and their barred counterparts are rational functions (again, up to integer powers of $\sqrt{\gamma^{2}-1}$).
For the full expression we refer the reader to the ancillary file.

As checks on these two observables we have confirmed:
(i) the cancellation of all $1/\eps$ poles occurring between the (PP) and (RR) regions;
(ii) conservation of $p_i^2$, $S_{i}^{2}$ and the $\cN=1$ global supercharge $Q_i=p_i\cdot\psi_i$.
While the first two  only check the simpler terms  carrying $F_\alpha^{(v)}$,
the latter also compares $F_\alpha^{(b)}$ terms between 
$\Delta p_{i,\rm cons}^{(4)\mu}$ and  $\Delta\psi_{i,\rm cons}^{(4)\mu}$, and thus $\Delta S_{i,\rm cons}^{(4)\mu}$.

We also define the total scattering angle $\theta$ for generic spin configurations as
\begin{align}
   \sin\frac\theta2
   =
   \frac{|\Delta p^\mu_{i,\rm cons}|}{2 p_\infty},\quad
   \theta= \frac{E}{M} \sum_{n,m}\left(\frac{GM}{|b|}\right)^n\frac{\theta^{(n,m)}}{|b|^m}\,,
\end{align}
with $p_\infty=m_1 m_2 \sqrt{\gamma^2-1}/E$, total energy ${E=|p_1^\mu+p_2^\mu|}$
and total mass $M=m_1+m_2$, $n$ and $m$ counting PM and spin orders respectively.
The 4PM spin-orbit contribution is
 \begin{align}
\begin{aligned}
 \label{scatang}
 \theta^{(4,1)}_{\text{cons}}=
\sum_{\alpha=1}^{16}\pi \nu 
 \bigg (
 s_{+} h^{(+)}_{\alpha}(\gamma) + \delta \, s_{-} h^{(-)}_{\alpha}(\gamma)
 \bigg)\, F^{(b)}_{\alpha}(\gamma)&\\ 
 -\frac{21 \pi  \gamma  \left(33 \gamma ^4-30 \gamma ^2+5\right) (13 s_{+}-3 \delta 
   s_{-})}{32 \left(\gamma ^2-1\right)^{5/2}}\,,&
\end{aligned}
\end{align}
where the test-body contributions (second line) agree with the geodesic motion in a Kerr background~\cite{Damgaard:2022jem}.
Here we use the mass parameters $\nu=m_1 m_2/M^2$ and $\delta=(m_2-m_1)/M$ and we have defined $s_\pm = -(a_1\pm a_2)\cdot \hat L$.
The 32 polynomial functions $h^{(\pm)}_{\alpha}(\gamma)$ are given in the
supplementary material~\eqref{polys}.
We have checked this result against the corresponding 
N${}^{3}$LO PN~\cite{Mandal:2022nty,Kim:2022pou}
literature, and found agreement by taking the PN expansion.
The tail term $P^{(4)}_\theta(\gamma)$ of the scattering angle is simply related to the 3PM radiated energy $E_{\rm rad}^{(3)}$ as follows:
% \begin{align}
%   P_\theta(\gamma)
%   =
%   G^4
%   \frac{\Gamma^2}{\sqrt{\gamma^2-1} \nu}
%   \frac{
%     \pat E_{\rm rad}^{(3)}}{
%     \pat |b|}
% \end{align}
\begin{align}
  P^{(4)}_\theta(\gamma)
  =
  E
  \frac{
    \pat E_{\rm rad}^{(3)}}{
    \pat J}\,,
\end{align}
where $J=p_{\infty}|b|$ is the initial angular momentum.
This equation follows the pattern derived in Ref.~\cite{Bini:2017wfr} and constitutes another non-trivial check of our results.
All of our results are included in an ancillary file attached to the \texttt{arXiv}
submission of this Letter.

\sec{Outlook} 
Having produced a complete set of 4PM linear-in-spin conservative scattering observables ---
and successfully compared them with N$^3$LO spin-orbit PN~\cite{Mandal:2022nty,Kim:2022pou}---
our next step will be upgrading them to include dissipative effects,
as has already been done in the non-spinning case~\cite{Dlapa:2022lmu,Dlapa:2023hsl}.
This will require two changes to our setup:
retarded graviton propagators in place of time-symmetric Feynman (see \Rcite{Jakobsen:2022psy})
and incorporation of the (PR+RP) regions when fixing boundary conditions on master integrals.
Notwithstanding the added complexity, quadratic-in-spin order is also an achievable target ---
corresponding N$^3$LO quadratic-in-spin PN results are already available~\cite{Mandal:2022ufb,Kim:2022bwv}.
In the near future we also seek to use these results to describe bound orbits,
the main obstacle being the aforementioned tail effect~\cite{Bern:2021yeh,Dlapa:2021vgp}.
Recent Numerical Relativity simulations of spinning black holes on hyperbolic-like orbits
\cite{Hopper:2022rwo} also offer us future numerical comparisons of the scattering angle $\theta$.

\sec{Acknowledgments}
We thank Alessandra Buonanno, Christoph Dlapa, Gregor K\"alin, Jung-Wook Kim, Zhengwen Liu, 
Raj Patil, Chia-Hsien Shen and
Jan Steinhoff for enlightening discussions and Peter Uwer for help with high-performance computing.
This work is funded by the Deutsche Forschungsgemeinschaft
(DFG, German Research Foundation)
Projektnummer 417533893/GRK2575 ``Rethinking Quantum Field Theory''.

\newpage
\allowdisplaybreaks

\appendix
\begin{widetext}
\section*{Supplementary Material}
\sec{Scattering angle} 
The 32 rational functions $h^{\pm}_{\alpha}(\gamma)$ 
(up to integer powers of $\sqrt{\gamma^{2}-1}$) appearing
in the spin-orbit contribution to the scattering angle~\eqref{scatang} take the explicit form
\begin{footnotesize}
\begin{align}
\begin{split}
h^{(+)}_{1}&= 
\frac{3 \pi ^2 (\gamma +1)^2 (1225 \gamma ^8+1225 \gamma ^7-1875 \gamma ^6-1875 \gamma ^5+795 \gamma ^4+3035 \gamma ^3-3601 \gamma ^2+1775 \gamma -448 )}
{192 (\gamma +1)^3 \sqrt{\gamma ^2-1}}
-\frac{1}{192 \gamma ^8 (\gamma +1) (\gamma ^2-1)^{5/2}}
\bigg (
22050 \gamma ^{19}
\\&
+33075 \gamma ^{18}-71725 \gamma ^{17}-123397 \gamma ^{16}+186555 \gamma ^{15}+67503 \gamma ^{14}-89885 \gamma ^{13}-190167 \gamma ^{12}+181103 \gamma ^{11}+137042 \gamma ^{10}-506830 \gamma ^9\\&
+407004 \gamma ^8 -33671 \gamma ^7-33671 \gamma ^6+8501 \gamma ^5+8501 \gamma ^4-1885 \gamma ^3-1885 \gamma ^2+315 \gamma +315
\bigg )
\\
h^{(-)}_{1}&=
\frac{3 \pi ^2 (-1225 \gamma ^8-1225 \gamma ^7+1875 \gamma ^6+1875 \gamma ^5-795 \gamma ^4-1115 \gamma ^3+401 \gamma ^2-111 \gamma +64) (\gamma +1)^2}
{192 (\gamma +1)^3 \sqrt{\gamma ^2-1}}
+\frac{1}{192 \gamma ^8 (\gamma +1) (\gamma ^2-1)^{5/2}}
\bigg (
22050 \gamma ^{19} \\&
+33075 \gamma ^{18}-71725 \gamma ^{17}-115333 \gamma ^{16}+96699 \gamma ^{15}+140871 \gamma ^{14}-56261 \gamma ^{13}-73191 \gamma ^{12}-6593 \gamma ^{11}+27498 \gamma ^{10}-3718 \gamma ^9
\\&
+9004 \gamma ^8-1491 \gamma ^7
-1491 \gamma ^6+313 \gamma ^5+313 \gamma ^4+95 \gamma ^3+95 \gamma ^2-105 \gamma -105
\Bigg )
\\  
h^{(+)}_{2}&=
 -\frac{    \gamma  \left(2 \gamma ^2-3\right) \left(420 \gamma ^6-1180 \gamma ^5-875 \gamma ^4-2587 \gamma ^3+10515 \gamma ^2-8829 \gamma +2752\right)}{8 (\gamma -1)^3 (\gamma +1)^4}\\
h^{(-)}_{2}&=
-\frac{    \gamma  \left(2 \gamma ^2-3\right) \left(160 \gamma ^5-25 \gamma ^4+199 \gamma ^3-739 \gamma ^2+493 \gamma -112\right)}{8 (\gamma -1)^3 (\gamma +1)^4}\\
h^{(+)}_{3}&=
 \frac{    -3675 \gamma ^{10}+11750 \gamma ^8-13720 \gamma ^6-10342 \gamma ^4-74445 \gamma ^2-11200}{16 \left(\gamma ^2-1\right)^{5/2}}\quad
h^{(-)}_{3}=
\frac{3675 \gamma ^{10}-11750 \gamma ^8+13720 \gamma ^6-5018 \gamma ^4+5837 \gamma ^2+448}{16 \left(\gamma ^2-1\right)^{5/2}}\\
h^{(+)}_{4}&=
 \frac{3675 \gamma ^{10}-11750 \gamma ^8+1680 \gamma ^7+17940 \gamma ^6+27396 \gamma ^5+20930 \gamma ^4+85992 \gamma ^3+150881 \gamma ^2+71748 \gamma +21588}{32 \left(\gamma ^2-1\right)^{5/2}}\\
h^{(-)}_{4}&=
\frac{-3675 \gamma ^{10}+11750 \gamma ^8-14028 \gamma ^6-2724 \gamma ^5+3478 \gamma ^4-6872 \gamma ^3-10473 \gamma ^2-2884 \gamma -876}{32 \left(\gamma ^2-1\right)^{5/2}}\\
h^{(+)}_{5}&=
 \frac{420 \gamma ^6-635 \gamma ^5-330 \gamma ^4-2977 \gamma ^3+10125 \gamma ^2-8984 \gamma +2597}{8 (\gamma -1) (\gamma +1)^2 \sqrt{\gamma ^2-1}}
\quad
h^{(-)}_{5}=
\frac{77 \gamma ^5-108 \gamma ^4+277 \gamma ^3-661 \gamma ^2+498 \gamma -107}{8 (\gamma -1) (\gamma +1)^2 \sqrt{\gamma ^2-1}}\\
h^{(+)}_{6}&= -9  h^{(-)}_{6}
=
-\frac{27 \gamma ^3 \left(3-2 \gamma ^2\right)^2 \left(7 \gamma ^2-3\right)}{16 \left(\gamma ^2-1\right)^{7/2}}
%\\
%
%h^{(-)}_{6}&=
%\frac{3 \pi  \gamma ^3 \left(3-2 \gamma ^2\right)^2 \left(7 \gamma ^2-3\right)}{16 %\left(\gamma ^2-1\right)^{7/2}}
%\\
\quad
h^{(+)}_{7}=-9h^{(-)}_{7}=\frac{27  \gamma ^2 \left(14 \gamma ^6-201 \gamma ^4+212 \gamma ^2+87\right)}{8 \left(\gamma ^2-1\right)^3}
%\\
%
%h^{(-)}_{7}&=-\frac{3 \pi  \gamma ^2 \left(14 \gamma ^6-201 \gamma ^4+212 \gamma ^2+87\right)}%{8 \left(\gamma ^2-1\right)^3}
%\\
\\
h^{(+)}_{8}&= -9 h^{(-)}_{8}=\frac{27 \gamma ^2 \left(14 \gamma ^4-27 \gamma ^2+9\right)}{8 \left(\gamma ^2-1\right)^2}
%\\
%
%h^{(-)}_{8}&=-\frac{3 \pi  \gamma ^2 \left(14 \gamma ^4-27 \gamma ^2+9\right)}{8 \left(\gamma %^2-1\right)^2}
\quad
h^{(+)}_{9}=-\frac{3   \left(63 \gamma ^4+203 \gamma ^3-247 \gamma ^2+89 \gamma -28\right)}{4 (\gamma +1) \sqrt{\gamma ^2-1}}
\quad
h^{(-)}_{9}=-\frac{3   \left(-7 \gamma ^4-27 \gamma ^3+23 \gamma ^2-9 \gamma +4\right)}{4 (\gamma +1) \sqrt{\gamma ^2-1}}
\\  
h^{(+)}_{10}&=\frac{3  \left(35 \gamma ^4+180 \gamma ^3+84 \gamma ^2+72 \gamma +7\right)}{\left(\gamma ^2-1\right)^{3/2}}
\quad
h^{(-)}_{10}=\frac{3  \left(-5 \gamma ^4-20 \gamma ^3-8 \gamma ^2-8 \gamma -1\right)}{\left(\gamma ^2-1\right)^{3/2}}
\\  
h^{(+)}_{11}&=\frac{3   \left(2240 \gamma ^6-3520 \gamma ^5-1504 \gamma ^4+4832 \gamma ^3-2784 \gamma ^2+672 \gamma \right)}{64 (\gamma -1)^2 (\gamma +1)^3}
\\&
+\frac{3   \left(-1225 \gamma ^{12}-1225 \gamma ^{11}+4325 \gamma ^{10}+4325 \gamma ^9-5770 \gamma ^8-3530 \gamma ^7+11546 \gamma ^6+12442 \gamma ^5-1277 \gamma ^4-9341 \gamma ^3-7151 \gamma ^2-2671 \gamma -448\right)}{64 (\gamma -1)^2 (\gamma +1)^3 \sqrt{\gamma ^2-1}}
\\
h^{(-)}_{11}&=\frac{3  \left(-320 \gamma ^6+320 \gamma ^5+288 \gamma ^4-416 \gamma ^3+288 \gamma ^2-96 \gamma \right)}{64 (\gamma -1)^2 (\gamma +1)^3}
\\&
+\frac{3  \left(1225 \gamma ^{12}+1225 \gamma ^{11}-4325 \gamma ^{10}-4325 \gamma ^9+5770 \gamma ^8+5450 \gamma ^7-4506 \gamma ^6-4378 \gamma ^5+1149 \gamma ^4+1789 \gamma ^3+623 \gamma ^2+239 \gamma +64\right)}{64 (\gamma -1)^2 (\gamma +1)^3 \sqrt{\gamma ^2-1}}
\\  
h^{(+)}_{12}&=\frac{3  \gamma  \left(1225 \gamma ^8-3100 \gamma ^6+2670 \gamma ^4+4884 \gamma ^2+2385\right)}{32 \left(\gamma ^2-1\right)^{3/2}}
\quad
h^{(-)}_{12}=\frac{3   \gamma  \left(-1225 \gamma ^8+3100 \gamma ^6-2670 \gamma ^4+236 \gamma ^2-337\right)}{32 \left(\gamma ^2-1\right)^{3/2}}
\\  
h^{(+)}_{13}&=-\frac{6   \gamma  \left(2 \gamma ^2-3\right) \left(35 \gamma ^3-55 \gamma ^2+29 \gamma -7\right)}{(\gamma -1)^2 (\gamma +1)^3}
\quad
h^{(-)}_{13}=-\frac{6  \gamma  \left(2 \gamma ^2-3\right) \left(-5 \gamma ^3+5 \gamma ^2-3 \gamma +1\right)}{(\gamma -1)^2 (\gamma +1)^3}
\\  
h^{(+)}_{14}&=\frac{120 \gamma ^4+13740 \gamma ^3+31705 \gamma ^2+22164 \gamma +5167}{8 \left(\gamma ^2-1\right)^{5/2}}
\quad
h^{(-)}_{14}=\frac{120 \gamma ^4-660 \gamma ^3-2515 \gamma ^2-2468 \gamma -733}{8 \left(\gamma ^2-1\right)^{5/2}}
\\  
h^{(+)}_{15}&=\frac{5  \gamma  \left(24 \gamma ^4+3509 \gamma ^2+3539\right)}{8 (\gamma -1)^2 (\gamma +1) \sqrt{\gamma ^2-1}}
\quad
h^{(-)}_{15}=\frac{  \gamma  \left(120 \gamma ^4-1135 \gamma ^2-2113\right)}{8 (\gamma -1)^2 (\gamma +1) \sqrt{\gamma ^2-1}}
\\  
h^{(+)}_{16}&=\frac{-240 \gamma ^5-13860 \gamma ^4-35090 \gamma ^3-53869 \gamma ^2-35390 \gamma -5167}{8 \left(\gamma ^2-1\right)^{5/2}}
\quad
h^{(-)}_{16}=\frac{-240 \gamma ^5+540 \gamma ^4+2270 \gamma ^3+4983 \gamma ^2+4226 \gamma +733}{8 \left(\gamma ^2-1\right)^{5/2}}
\end{split}\label{polys}
\end{align}
\end{footnotesize}
Here the $(\pm)$ upper indices label the coupling to the spin-orbit 
components  $s_\pm = -(a_1\pm a_2)\cdot \hat L$ via $s_{+}$ to $h_{\alpha}^{(+)}$ and
$\delta \, s_{-}$ to $h_{\alpha}^{(-)}$.
%The indices $\alpha=1,\ldots, 16$ correspond to the entries 
%of the function space vector $F^{(b)}_{\alpha}$,
%i.e.
%%
%\begin{align}
%F^{(b)}_{\alpha}(\gamma)&=\{1,\text{arccosh}[\gamma], \log [\gamma], \log \left[\frac{\gamma_{+}}{2} 
% \right], \log \left[\frac{\gamma_{-}}{2} 
% \right],
% \text{arccosh}^{2}[\gamma], 
% \text{arccosh}[\gamma] \log \left[\frac{\gamma_{+}}{2} \right ],
% \text{arccosh}[\gamma] \log \left[\frac{\gamma_{-}}{2} \right ],\\ &
%  \log \left[\frac{\gamma_{+}}{2} \right ]\log \left[\frac{\gamma_{-}}{2} \right ],
%   \log^{2} \left[\frac{\gamma_{+}}{2}\right ],
%    \text{Li}_{2}\left [\frac{\gamma_{-}}{\gamma_{+}}\right ], 
%     \text{Li}_{2}\left [-\frac{\gamma_{-}}{\gamma_{+}}\right],  
% \text{Li}_{2}\left [\sqrt{\frac{\gamma_{-}}{\gamma_{+}}}\right], 
%   \eK^2\left [ \frac{\gamma_{-}}{\gamma_{+}} \right ],
%   \eE^2\left [ \frac{\gamma_{-}}{\gamma_{+}} \right ],
%   \eK\left [ \frac{\gamma_{-}}{\gamma_{+}} \right ]
%   \eE\left [ \frac{\gamma_{-}}{\gamma_{+}} \right ]
%   \}\,,\nn
%\end{align}
%which provides a precise ordering for the list in \Eqn{eq:functions}.
The indices $\alpha=1,\ldots, 16$ correspond to the entries 
of the function space vector $F^{(b)}_{\alpha}$ of \Eqn{eq:functions}
where we follow the convention that (+) entries come before (-) entries when a $(\pm)$ appears.
%The indices $\alpha=1,\ldots, 16$ correspond to the entries of the function space vector $F^{(b)}_{\alpha}$ of \Eqn{eq:functions} where we follow the convention that $\gamma_{+}$ entries come before $\gamma_{-}$ ones.

\end{widetext}

\bibliographystyle{JHEP}
\bibliography{spin-4PM}

\end{document}